\documentclass[conference]{IEEEtran}
 
\usepackage[utf8]{inputenc}
\usepackage{kotex}
\usepackage{booktabs}
\usepackage{caption}
\usepackage{subcaption}
\usepackage{multirow}
\usepackage[hang]{footmisc}
\usepackage{hyperref}

\usepackage{cite}
\usepackage{amsmath,amssymb,amsfonts}
\usepackage{algorithmic}
\usepackage{graphicx}
\usepackage{textcomp}
\usepackage{xcolor}
\def\BibTeX{{\rm B\kern-.05em{\sc i\kern-.025em b}\kern-.08em
    T\kern-.1667em\lower.7ex\hbox{E}\kern-.125emX}}

\usepackage{listings}
\definecolor{codegreen}{rgb}{0,0.6,0}
\definecolor{codegray}{rgb}{0.5,0.5,0.5}
\definecolor{codepurple}{rgb}{0.58,0,0.82}
\definecolor{backcolour}{rgb}{0.95,0.95,0.92}

\lstdefinestyle{mystyle}{
    backgroundcolor=\color{backcolour},   
    commentstyle=\color{codegreen},
    keywordstyle=\color{magenta},
    numberstyle=\tiny\color{codegray},
    stringstyle=\color{codepurple},
    basicstyle=\ttfamily\scriptsize,
    breakatwhitespace=false,         
    breaklines=true,                 
    captionpos=b,                    
    keepspaces=true,                 
    numbers=left,                    
    numbersep=5pt,                  
    showspaces=false,                
    showstringspaces=false,
    showtabs=false,                  
    tabsize=2
}

\begin{document}

\title{Single-Channel Distance-Based Source Separation for Mobile GPU in Outdoor and Indoor Environments}

\author{\IEEEauthorblockN{1\textsuperscript{st} Hanbin Bae$^*$}
\IEEEauthorblockA{\textit{AI Solution Team} \\
\textit{Samsung Research}\\
Seoul, Republic of Korea \\
bhb0722.bae@samsung.com}
\and
\IEEEauthorblockN{2\textsuperscript{nd} Byungjun Kang$^*$}
\IEEEauthorblockA{\textit{AI Solution Team} \\
\textit{Samsung Research}\\
Seoul, Republic of Korea \\
bj88.kang@samsung.com}
\and
\IEEEauthorblockN{3\textsuperscript{rd} Jiwon Kim}
\IEEEauthorblockA{\textit{AI Solution Team} \\
\textit{Samsung Research}\\
Seoul, Republic of Korea \\
jiwon177.kim@samsung.com}
\and
\IEEEauthorblockN{4\textsuperscript{th} Jae-Yong Hwang}
\IEEEauthorblockA{\textit{AI Solution Team} \\
\textit{Samsung Research}\\
Seoul, Republic of Korea \\
j\_yong.hwang@samsung.com}
\and
\IEEEauthorblockN{5\textsuperscript{th} Hosang Sung}
\IEEEauthorblockA{\textit{AI Solution Team} \\
\textit{Samsung Research}\\
Seoul, Republic of Korea \\
hosang.sung@samsung.com}
\and
\IEEEauthorblockN{6\textsuperscript{th} Hoon-Young Cho}
\IEEEauthorblockA{\textit{AI Solution Team} \\
\textit{Samsung Research}\\
Seoul, Republic of Korea \\
h.y.cho@samsung.com}
}

\newcommand{\red}[1]{\textcolor{red}{#1}}

\maketitle
\def\thefootnote{*}\footnotetext{The authors contributed equally to this work}
\begin{abstract}

This study emphasizes the significance of exploring distance-based source separation (DSS) in outdoor environments. Unlike existing studies that primarily focus on indoor settings, the proposed model is designed to capture the unique characteristics of outdoor audio sources. It incorporates advanced techniques, including a two-stage conformer block, a linear relation-aware self-attention (RSA), and a TensorFlow Lite GPU delegate. 
While the linear RSA may not capture physical cues as explicitly as the quadratic RSA, the linear RSA enhances the model's context awareness, leading to improved performance on the DSS that requires an understanding of physical cues in outdoor and indoor environments.
The experimental results demonstrated that the proposed model overcomes the limitations of existing approaches and considerably enhances energy efficiency and real-time inference speed on mobile devices.

\end{abstract}


\section{Introduction}

The recently developed distance-based source separation (DSS) \cite{patterson22_interspeech, lin23b_interspeech, petermann2024hyperbolic} separates a  single-channel audio mixture into individual components based on their distance from a reference point, such as a microphone. The sources are classified as ’near’ or ’far’ based on a specified distance threshold. In contrast to traditional methods, this method emphasizes upon the spatial information of the sources, particularly their distance from the recording device, to improve the separation process in challenging acoustic scenarios. For example, in environments with multiple sound sources at different distances from a microphone, (i.e., conference rooms, lecture halls, and concert venues), DSS facilitates the isolation of individual speakers.

To effectively use a DSS model in a single-channel setting, the physical information must be appropriately interpreted to better perceive the distance. This involves factors such as the intensity variation with distance (based on the inverse-square law), direct-to-reverberation ratio (DRR), proximity effects that result in directional microphones emphasizing low frequencies for sources that are closer, and acoustic cues such as absorption by air and spatial effects \cite{patterson22_interspeech}. Based on the understanding and exploitation of these physical cues, a DSS model can effectively distinguish between near and far sound sources considering their distance from the microphone, thereby enabling accurate sound separation. 

However, recent studies \cite{patterson22_interspeech, lin23b_interspeech, petermann2024hyperbolic} have solely focused on sound separation in indoor environments; despite the necessity of investigating its application for outdoor environments. In outdoor environments with high noise, poor performance is exacerbated by the greater impact of the difficulty of measuring DRR. Further, environmental factors, such as wind, ambient noise, and varying distances of sound sources can hinder the accurate separation of near and far sounds. In addition, the presence of background noise and unpredictable acoustic conditions outdoors can degrade the performance of separation models trained on indoor data. The application of DSS models to outdoor environments requires robust algorithms that can incorporate these environmental variables.

The requirement of additional support of a DSS technique on mobile devices is evident when examining how people aim to improve audio for their self-created videos. For instance, vlogs, wherein users record and share their daily experiences, are among the most popular types of content, and most are created using mobile phones. Prior to conducting this study, we conducted a preliminary survey with \textit{amateur experts}~\cite{Kuznetsov:2010} aged 20-30 who frequently shot and uploaded videos on social media. This study revealed that one of the biggest challenges faced by the participants was the removal of unintentional sounds from their video recordings. 
These videos typically have a running time of 5 minutes and may be longer than an hour. Consequently, the inference speed and resource efficiency of the model should be maximized.

This study proposes a DSS model architecture and training method that could support mobile graphical processing units (GPUs) and provide robust performance in outdoor and indoor environments. The proposed architecture utilizes the EfficientAttention \cite{shen2021efficient} mechanism and the Rotary Positional Embedding (RoPE) \cite{su2021rope} to guarantee a 20 times increase in energy efficiency and a 10 times faster inference speed compared to mobile central processing units (CPUs) without the loss of separation quality.
To the best of our knowledge, this is the first study involving the application of DSS to a noisy outdoor environment.

\section{Problem Formulation}

In this study, we assume that the input mixture signal $x^{mix}$ is approximated as follows:

\begin{equation} \label{eq:problem} 
\resizebox{.7\hsize}{!}{$
\begin{split}
x^{mix} & \approx \underbrace{ \sum^{N^{near}}_{i}x^{near}_{i}*h^{near}_{i} }_{y^{near}}+ \underbrace{ \sum^{N^{far}}_{j}x^{far}_{j}*h^{far}_{j}+\epsilon }_{y^{far}},
\end{split}
$}
\end{equation}
where $N$, $x$, $\epsilon$, and $h$ denote the number of sources, the direct path signals of each source, the background noise signal, and the impulse responses of each source in an acoustic environment.

To simulate the impulse responses of each source, the Pyroomacoustics \cite{scheibler2018pyroom} toolbox's Image Source Model (ISM) \cite{allen1979ism} was used. This model considered all possible sound wave paths from the source to the virtual sources and then to the microphone, capturing the spatial and reverberation characteristics of the room. 
We then convolved each source's direct-path signal and impulse response and added the background noise into the convolved signals.

\begin{figure}[t]
    \centering
    \centerline{\includegraphics[width=0.6\linewidth]{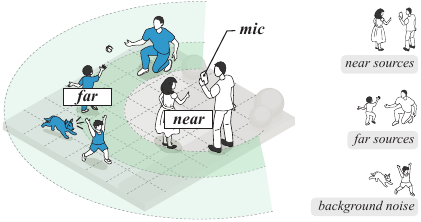}}
\caption{Visual representation of DSS with background noise.}
\label{fig_teaser}
\vspace{-0.5cm}
\end{figure}

With the increase in the distance, the intensity of the direct path signal and the time difference between the direct path signal and the early reflection signals reduces. Consequently, distinguishing the acoustic characteristics of the far signals from background noise, particularly babble noise, is challenging. Therefore, we did not distinguish them and defined the first term of the right-hand side of eq~(\ref{eq:problem}) as $y^{near}$, and the last two terms were combined as $y^{far}$; this is visually represented in Fig.~\ref{fig_teaser}.

\section{Proposed Architecture} \label{sec:proposed_architecture}

\subsection{Baseline Architecture}
\begin{figure}[t]
    \centering
    \centerline{\includegraphics[width=0.9\linewidth]{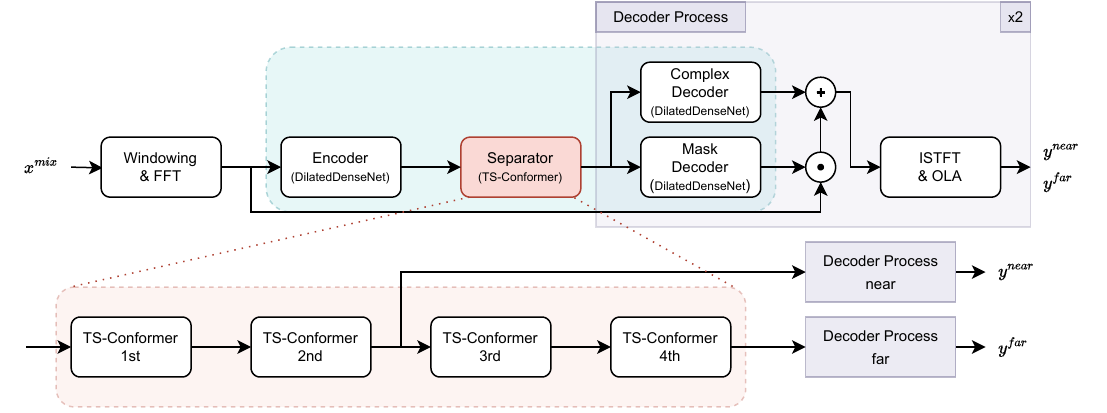}}
\caption{Schematic diagrams of $\textrm{M}_{\textrm{Baseline}}$  architecture.}
\label{fig_baseline_architecture}
\vspace{-0.5cm}
\end{figure}

\label{sec
} We adopt a conformer-based metric generative adversarial network (CMGAN)\cite{cao2022cmgan} as the baseline architecture $\textrm{M}_{\textrm{Baseline}}$, shown in Fig.~\ref{fig_baseline_architecture}. CMGAN combines conformer blocks~\cite{gulati2020conformer} in a dual-path design and metric GAN~\cite{fu2019metricgan} for optimizing speech enhancement performance. However, the metric GAN is excluded as designing its metric and discriminator is beyond this study's scope.

\textbf{Encoder Process:} The input waveform is transformed into a complex spectrogram via STFT (FFT size: 512, hop size: 128, Hamming window) and compressed using a power-law technique~\cite{braun2020powerlaw} with a 0.3 exponent. The encoded spectrogram passes through a dilated DenseNet~\cite{pandey2020densenet} with four convolution blocks (dilations: {1, 2, 4, 8}), halving the frequency dimension in the last block.

The separator uses four two-stage conformer (TS-Conformer) blocks to capture time and frequency dependencies via multi-head self-attention (MHSA)~\cite{vaswani2017transformer} and CNN layers. The outputs of the 2nd and 4th blocks are then passed to the decoder for near and far targets, respectively.

\textbf{Decoder Process:} The decoder consists of two paths: Mask Decoder and Complex Decoder, both using dilated DenseNet and sub-pixel convolution for frequency upsampling. The Mask Decoder predicts the spectral mask, while the Complex Decoder predicts the real and imaginary parts of the spectrogram. The final spectrogram is obtained by multiplying the mixture spectrogram by the mask and adding the complex spectrogram. It then undergoes inverse power-law compression and ISTFT to produce the time-domain signal.

\subsection{Linearization of relation-aware self-attention}

The conformer architectures \cite{cao2022cmgan, chen2020continuous} use relation-aware self-attention (RSA) \cite{shaw2018selfattention} as the core mechanism in MHSA. RSA enhances traditional self-attention by considering both the absolute positions and the relative relationships between elements in a sequence, enabling the model to capture more nuanced interactions. This allows for more flexible and context-aware processing, improving performance in DSS models that require understanding time and frequency domain information. The RSA operation is represented as:

\begin{equation} \label{eq:einsum}
\resizebox{.7\hsize}{!}{$
\begin{aligned}
Output &= \textrm{softmax} ((QK^{T} + Q \odot R )/ \sqrt{d})V \\
Q \odot R &= \textrm{einsum}(\textrm{`B N d, N M d} \rightarrow \textrm{B N M'}, Q, R),
\end{aligned}
$}
\end{equation}
where $K$, $Q$, $V$, and $R$ represent key, query, value, and relative-positional embedding matrices, while B, N, M, and d denote batch size, sequence length, alternate sequence length for einsum, and dimensionality, respectively.

In eq~(\ref{eq:einsum}), this attention mechanism leads to $O(N^{2}d)$ calculations, and as the sequence length increases, the complexity grows quadratically. This issue has been widely discussed in recent papers~\cite{child2019generating,yue2018compact,zhang2019latentgnn,zhang2021linearspeech,shen2021efficient} and causes the inefficient energy consumption and slow inference speed. 

To address this, we adopt a linear RSA inspired by LinearSpeech \cite{zhang2021linearspeech}, combining EfficientAttention and RoPE. EfficientAttention simplifies self-attention with kernel feature maps and a linear dot product, while RoPE handles relative positional representations, enabling efficient long-sequence synthesis. The linear RSA is represented as:

\begin{equation} \label{eq:linearattn}
\resizebox{.5\hsize}{!}{$
\begin{aligned}
Output &= \sigma_{q}(Q_{r}) \{\sigma_{k}(K_{r}^{T})V\} / \sqrt{d},
\end{aligned}
$}
\end{equation}
where $Q_{r} = f_{\textrm{RoPE}}(Q)$, $K_{r}=f_{\textrm{RoPE}}(K)$, and $\sigma_{k}$ and $\sigma_{q}$ denote kernel feature maps for key and query, and $f_{\textrm{RoPE}}$ denotes the function of RoPE. In this study, the softmax function is used as both kernel feature maps. The RoPE does not use a trainable parameter, whereas the $R$ uses multiple trainable embedding parameters. This linear RSA contributes to the model's efficiency in memory, fast convergence of training, and computational complexity  $O(Nd^{2})$. 

\subsection{Mobile GPU Utilization}

Mobile GPUs are optimized for parallel processing, making them ideal for tasks like deep neural network inference. They provide advantages over mobile CPUs, such as higher parallelism, faster execution of complex computations, potential real-time performance, and improved energy efficiency. Integrating the GPU delegate in TensorFlow Lite (TFLite) enhances model efficiency, resulting in lower latency.

The following two rules should be considered to ensure a GPU delegate works with every operation.
\begin{enumerate}
    \item Use layers supported by the GPU delegate.
    \item Maintain a consistent batch size for all tensors.
\end{enumerate}

We excluded reshape operations in TS-Conformer \cite{cao2022cmgan} to avoid changing tensor batch sizes and implemented a custom Conv1D layer to bypass TensorFlow's reshape operations. Additionally, we optimized all tensor shapes to be consistent (size 4) and adjusted the MHSA mechanism to handle input tensors separately along the h-axis. The BatchMatMul operation was optimized by disabling ``unfold\_batch\_matmul.''

\subsection{Benchmark Test of Architectures on Mobile Device} 
We evaluated the number of parameters, multiply-accumulate (MAC) operations, energy consumption, and inference real-time factor (RTF) on mobile CPUs and GPUs. Testing was conducted on a Galaxy S23 with battery protection at 85\%, and battery consumption was tracked using the dumpsys tool~\cite{dumpsys}. A 5-minute sample was played with intervals to avoid throttling, and input samples were segmented into 3-second chunks.

We also tested $\textrm{M}_{\textrm{Enc\&Dec}}$ (skipping all TS-Conformer blocks) and $\textrm{M}_{\textrm{TF-DPRNN}}$, which replaces TS-Conformer blocks with dual-path recurrent neural networks (DPRNN)~\cite{luo2020dprnn}.

Despite many efforts, $\textrm{M}_{\textrm{Baseline}}$ and $\textrm{M}_{\textrm{TF-DPRNN}}$ could not use the GPU delegate due to unsupported RSA and bi-directional LSTM layers. 

TABLE~\ref{tab:mobile_performance} summarizes the results. The $\textrm{M}_{\textrm{Proposed}}$ worked successfully with GPU delegate support, reducing energy consumption and inference speed by 20x and 10x compared to CPU. The TS-Conformer blocks with linear RSA were especially efficient on Mobile GPU, showing a 0.01\% energy use and 0.06 RTF, compared to 2.15\% energy use and 1.32 RTF on Mobile CPU.

\begin{table}[!h]\scriptsize
  % \vspace{-0.2cm}
    \centering
    \caption{Mobile benchmark performances of all architectures.}
      % \vspace{-0.1cm}
	\label{tab:mobile_performance}
	\begin{tabular}{l c c c c}
		\toprule
		\multirow{2}{*}{{\shortstack[c]{Model }}} & \multirow{2}{*}{{\shortstack[c]{\# Par. \\(M)}}}& \multirow{2}{*}{{\shortstack[c]{\# MAC/s \\ (G/s)}}} &    \multirow{2}{*}{{\shortstack[c]{CPU \\E (\%) / RTF } }} &  \multirow{2}{*}{{\shortstack[c]{GPU \\E (\%) / RTF } }}  \\
        &&&\\
  \midrule
        $\textrm{M}_{\textrm{Baseline}}$ &1.4& 29.5  & 4.85 / 2.01 & - \\
        $\textrm{M}_{\textrm{Proposed}}$ & 1.3 & 25.7 & 3.00 / 1.55 & 0.15 / 0.16  \\
  \midrule
        $\textrm{M}_{\textrm{Enc\&Dec}}$ &0.8& 14.5 & 0.85 / 0.27 & 0.14 / 0.10  \\
        $\textrm{M}_{\textrm{TF-DPRNN}}$ & 1.3 & 15.7 & 1.84 / 1.10 & - \\
  \bottomrule
	\end{tabular}
	 \vspace{-0.3cm}
\end{table}

\section{Experiments}

\subsection{Simulation outdoor and indoor environments}

To simulate the eq~(\ref{eq:problem}), we set ISM variables as follows: 
\begin{enumerate}
    \item The room dimensions (length, width, height) are randomly selected in $\left(6.0 \pm 1.5, 6.0 \pm 1.5, 2.6 \pm 0.2 \right)$ m. 
    \item The position of microphone is randomly selected in  (3.0 $\pm$ 0.7, 3.0 $\pm$ 0.7, 0.8 $\pm$ 0.7) m.
    \item The RT60s are randomly selected in (0.15, 1.0) sec. The distance threshold is set to $0.5$ m.
    \item The materials of all six walls are randomly selected from the toolbox in possible choices; all wall materials except the floor are set to `anechoic` for outdoor environments.
    \item The positions of near and far sources are randomly away from (0.02, 0.5) m and (1.3, 1.7) m, respectively. The far region not included in the training dataset is defined as the unseen region (UR).
\end{enumerate}
After applying impulse responses to each near and far source, the background noise was mixed into this signal within \{0, 5, 10, 15, 20\} dB signal-to-noise ratios (SNRs). All random variables follow the uniform distributions.

\subsection{Datasets}
For the training dataset, we used the CSTR-VCTK \cite{yamagishi2017vctk}, AISHELL \cite{shi2010aishell3}, AI-Hub datasets\footnote{This paper used datasets from `The Open AI Dataset Project (AI-Hub, S. Korea)'. All data information can be accessed through `AI-Hub (www.aihub.or.kr)'} \cite{aihub2024emotion,aihub2024translation,aihub2024noise} for source speakers and background noises.
The dataset contains approximately 5,000 hours of speech data, including about 500 speakers, 5 languages, and 8 emotion styles. It also contains approximately 1,000 hours of sound data from noisy environments, classified into various categories. The training dataset is divided into a training set and a validation set with a ratio of 9:1.
For the test dataset, we used the wsj0 \cite{garofolo2007wsj} for source speakers. We used the internal noise dataset that contains 30 hours of sound data from outdoor and indoor environments, classified into various categories. The training and test datasets were resampled at $16$ kHz sampling rate. 
The simulated datasets for training and test created 3-second and 5-second segments, respectively.

\subsection{Training setup}

Given the different characteristics of the outdoor and indoor samples, we conducted preliminary experiments on what ratios to use and how to train from both. The results showed that 60:40 was better than other ratios, and end-to-end training was better than fine-tuning training.

We set the hyperparameters related to all architectures, such as the number of two-stage conformer blocks, batch size, and channel size, to 4, 4, and 48, respectively, and the channel size and number of heads of MHSA to 48 and 4, respectively.

The AdamW optimizer \cite{loshchilov2019adamw} and the exponential learning rate scheduler for every 1,000 steps were used for training. The parameters of optimization, including the learning rate, beta1, beta2, epsilon, and learning rate decay, were set to 0.005, 0.8, 0.99, 1e-8, and 0.999. All models were updated by optimizing the weighted sum of the mag, spec, and time losses defined in CMGAN \cite{cao2022cmgan} with 0.9, 0.1, and 0.2 weights. 

\subsection{Evaluation of the basic performances of DSS models in indoor environments}

We tested for the simulated a total of 100 indoor samples using the test dataset and measured the average scale-invariant SDR improvement (SI-SDRi) \cite{roux2018sisdr} to confirm the separation performance. The results are shown in TABLE~\ref{tab:exp_indoor}. When neither near nor far source was included in the simulated sample, all models predominantly predicted silence. When we used mixed cases, $\textrm{M}_{\textrm{Proposed}}$ performed better than $\textrm{M}_{\textrm{Baseline}}$, indicating that the linear RSA was successfully replaced with. Moreover, we confirmed the importance of conformer blocks by observing the performance of $\textrm{M}_{\textrm{Roformer}}$, which replaces conformer blocks with transformer blocks using the RoPE\cite{su2021rope}.

The $\textrm{M}_{\textrm{TF-DPRNN}}$ exhibited superior performance when the target was near compared to the architectures using the TS-Conformer. Conversely, its performance was comparatively lower in the two cases where only one distance source existed. The near contains only speech, while the far contains background noise. The $\textrm{M}_{\textrm{TF-DPRNN}}$ performed well in separating only speech. If bidirectional LSTM is fully supported for TFLite GPU delegate, the DPRNN could be competitive for the DSS problem.

\begin{table}[!h]\scriptsize
  % \vspace{-0.3cm}
    \centering
    \caption{Test results of all architectures for indoor training and test datasets. Each cell represents the average SI-SDRi (dB) of (near / far).}
      % \vspace{-0.1cm}
	\label{tab:exp_indoor}
        \begin{tabular}{c c c c c}
		\toprule
		\ $\#n$ / $\#f$&  $\textrm{M}_{\textrm{TF-DPRNN}}$ &    $\textrm{M}_{\textrm{Baseline}}$ & $\textrm{M}_{\textrm{Roformer}}$ &$\textrm{M}_{\textrm{Proposed}}$\\
  \midrule
        0 / $\{1,2,3\}$ & 53.3 / - & 24.6 / - & 29.8 / - & 68.2 / -\\
        $\{1,2,3\}$ / 0 & - / 24.0 & - / 26.5 & - / 20.6 & - / 24.7\\
  \midrule
        1 / 1 & \textbf{6.0} / 11.7& 5.3 / 11.5 & 3.9 / 9.7 & 5.5 / \textbf{12.0}\\
        1 / 2& \textbf{7.0} / \textbf{9.9}& 5.9 / 9.5& 4.5 / 7.4 & 6.1 / 9.7\\
        2 / 1 & \textbf{2.7} / 9.0& 1.9 / 8.2& 1.5 / 7.7 & \textbf{2.7} / \textbf{9.4}\\
  \bottomrule
	\end{tabular}
	 \vspace{-0.1cm}
\end{table}

\subsection{Evaluation in outdoor environments}

\begin{table}[!h]\scriptsize	 
    \centering
    \caption{Test results of two $\textrm{M}_{\textrm{Proposed}}$ trained using the datasets mixed in 100:0 and 60:40 ratios, respectively. Each cell represents the average SI-SDRi (dB) of (near / far).}
      % \vspace{-0.1cm}
	\label{tab:exp_outdoor}
	\begin{tabular}{c c c c c c}
		\toprule
		 Env& I : O & UR-0 & UR-1  & SR& UR-2 \\
  \midrule
        \multirow{2}{*}{{\shortstack[c]{In}}}& 100 : 0 & 0.9 / 1.2 & 2.6 / 7.3& 3.3 / 9.1& 3.3 / 9.6\\
           & 60 : 40& \textbf{1.9} / \textbf{2.1}& \textbf{3.4} / \textbf{7.4}& \textbf{3.9} / \textbf{9.3}& \textbf{3.5} / \textbf{9.8}\\
  \midrule
        \multirow{2}{*}{{\shortstack[c]{Out}}}& 100 : 0 & 2.1 / -0.3 & 2.5 / 2.9& 3.6 / 6.9& 4.2 / 8.3\\
           & 60 : 40& \textbf{2.5} / \textbf{3.9}& \textbf{3.8} / \textbf{8.3}& \textbf{4.6} / \textbf{11.2}& \textbf{5.2} / \textbf{11.4}\\
         \bottomrule
	\end{tabular}
	 \vspace{-0.3cm}
\end{table}
\begin{figure}[t]
    \centering
    \centerline{\includegraphics[width=0.7\linewidth]{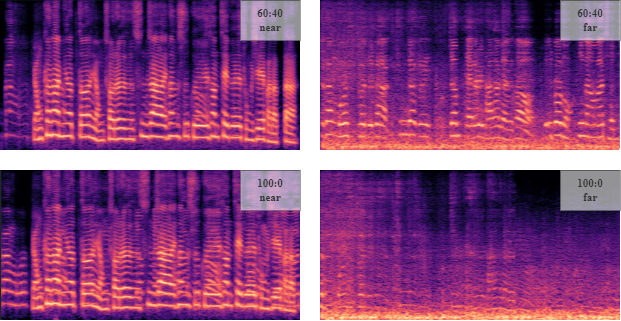}}
\caption{Results for a real outdoor sample. }
\label{fig:real_outdoor}
	 \vspace{-0.5cm}
\end{figure}
When simulating the far sources, we divided them into one seen region (SR) and three URs. UR-0, UR-1 and UR-2 represented the distance range of (0.5, 0.8) m, (0.8, 1.2) m, and (1.8, 2.2) m, respectively; we discussed the results of UR-0 in section~\ref{sec:limitation}.
We tested for the simulated a total of 200 outdoor and indoor samples with at least one speaker from both targets and measured the average SI-SDRi.
The results are shown in TABLE~\ref{tab:exp_outdoor}, and the output spectra of each model for the real outdoor sample are depicted in Fig.~\ref{fig:real_outdoor}, where the input sample is an audio sample of 5 seconds recording of two men speaking in a windy outdoor.
The overall performance of the proposed method, considering the outdoor dataset, was better than that considering the indoor dataset alone.

\subsection{Limitation and future work} \label{sec:limitation}
For the UR-0, we observed that some near sounds were frequently mixed in estimated far sounds, known as the permutation ambiguity problem \cite{liu2021pit}. 
Within this range, even humans have difficulty classifying an audio source as near and far recorded by a single-channel microphone.

To resolve the issue, ensuring that both outputs do not have the same source is crucial for improving the proposed model. A potential area for further exploration in this domain is to include embeddings that can hierarchically represent the distance and speaker information of sound sources  \cite{petermann2024hyperbolic}.

Audio samples can be found online\footnote{\url{https://icassp2025-hanbin-bae.netlify.app}}.

\section{Conclusion}
We present a new DSS model for mobile GPUs in outdoor and indoor environments. The proposed model uses a TS-Conformer block, linear complexity RSA, and TFLite GPU delegate to improve mobile device energy consumption and real-time inference speed. In both outdoor and indoor environment tests, the proposed model trained with a dataset that included outdoor environments outperformed the model trained only with indoor data.

\bibliographystyle{IEEEtran}
\bibliography{mybib}

\end{document}